\documentclass{XrU2005}

\usepackage{graphicx}

\title{Swift XRT Observations of X-ray Flares in GRB Afterglows}
\author[1]{David N. Burrows}
\author[2]{P. Romano}
\author[3]{O. Godet}
\author[1]{A. Falcone}
\author[1,2]{C. Pagani}
\author[4]{G. Cusumano}
\author[2]{S. Campana}
\author[2,5]{G. Chincarini}
\author[1,6]{J. E. Hill}
\author[7]{P. Giommi}
\author[3]{M. R. Goad}
\author[1]{J. A. Kennea}
\author[1,8] {S. Kobayashi}
\author[1]{P. M\'{e}sz\'{a}ros}
\author[1]{J. A. Nousek}
\author[3]{J. P. Osborne}
\author[3]{P. T. O'Brien}
\author[3]{K. L. Page}
\author[2]{G. Tagliaferri}
\author[9]{B. Zhang}
\author[10]{the Swift XRT team}
\affil[1]{Department of Astronomy \& Astrophysics, 525 Davey
  Lab., Penn. State University, University Park, PA 16802, USA}
\affil[2]{INAF-Osservatorio Astronomico di Brera, Via Bianchi 46, 23807 Merate, Italy}
\affil[3]{Department of Physics and Astronomy, University of Leicester, Leicester LE1 7RH, UK}
\affil[4]{INAF-Istituto di Astrofisica Spaziale e Fisica Cosmica,  
                 Via Ugo La Malfa 153, 90146 Palermo, Italy}
\affil[5]{Universit\`a degli studi di Milano-Bicocca,
                 Dipartimento di Fisica, Piazza delle Scienze 3, I-20126 Milan, Italy}
\affil[6]{NASA-Goddard Space Flight Center / Universities Space
  Research Assocation}
\affil[7]{ASI Science Data Center, via Galileo Galilei, 00044 Frascati,
  Italy}
\affil[8]{Astrophysics Research Institute, Liverpool John Moores
  University, Birkenhead CH41 1LD, UK}
\affil[9]{Department of Physics, University of Nevada, Box 454002, Las
  Vegas, NV, 89154-4002, USA}

\newcommand{\meszaros}{M\'{e}sz\'{a}ros}

\begin{document}

\keywords{GRBs; Swift; X-rays; afterglow}

\maketitle

\begin{abstract}
The Swift XRT has been observing GRB afterglows since December 23,
2004.  Three-quarters of these observations begin within 300 s of the
burst onset, providing an unprecendented look at the behavior of X-ray
emission from GRB afterglows in the first few hours after the burst.
While most of the early afterglows have smoothly declining
lightcurves, a substantial fraction has large X-ray flares on short
time-scales.  We suggest that these flares provide support
for models with extended central engine activity producing late-time internal shocks.
\end{abstract}

\section{Introduction}

The Swift Explorer mission \citep{Gehrels2004} was launched on November 20, 2004.  It is
detecting two bursts per week on average, and following these up with
detailed optical/UV and X-ray observations.  In ~75\% of the bursts,
the spacecraft can slew immediately to the field and 
observations with the X-ray Telescope \citep[XRT;][]{Burrows2005a}
begin within 5 minutes of the GRB trigger (in the
remaining cases the source is too close to the Earth, Moon, or Sun and
XRT observations are delayed).
While early X-ray observations are available for a handful of previous
bursts \citep[e.g., see][]{Piro2005}, the large number of these observations made
available by Swift is revolutionizing our knowledge base of early GRB
X-ray afterglows.  At the time of this writing (31 October 2005), the
XRT has detected 67 X-ray afterglows of GRBs (exceeding the total
pre-Swift afterglow sample), 51 of which were observed within 300 s of
the trigger.

Here we discuss the discovery of X-ray flares, commonly seen during
the first several hours after the burst.  These flares are seen in
approximately 50\% of all GRBs, and cover a range of time-scales
and intensities.  This paper will highlight some of the key findings
that led us to the conclusion that the flares are produced by
extended central engine activity producing X-rays from internal shocks
at times long after the cessation of hard X-ray/gamma-ray emission.

\section{GRB 050406}

\begin{figure*}
\centering
\includegraphics[width=5in,clip,bb=48 453 544 790]{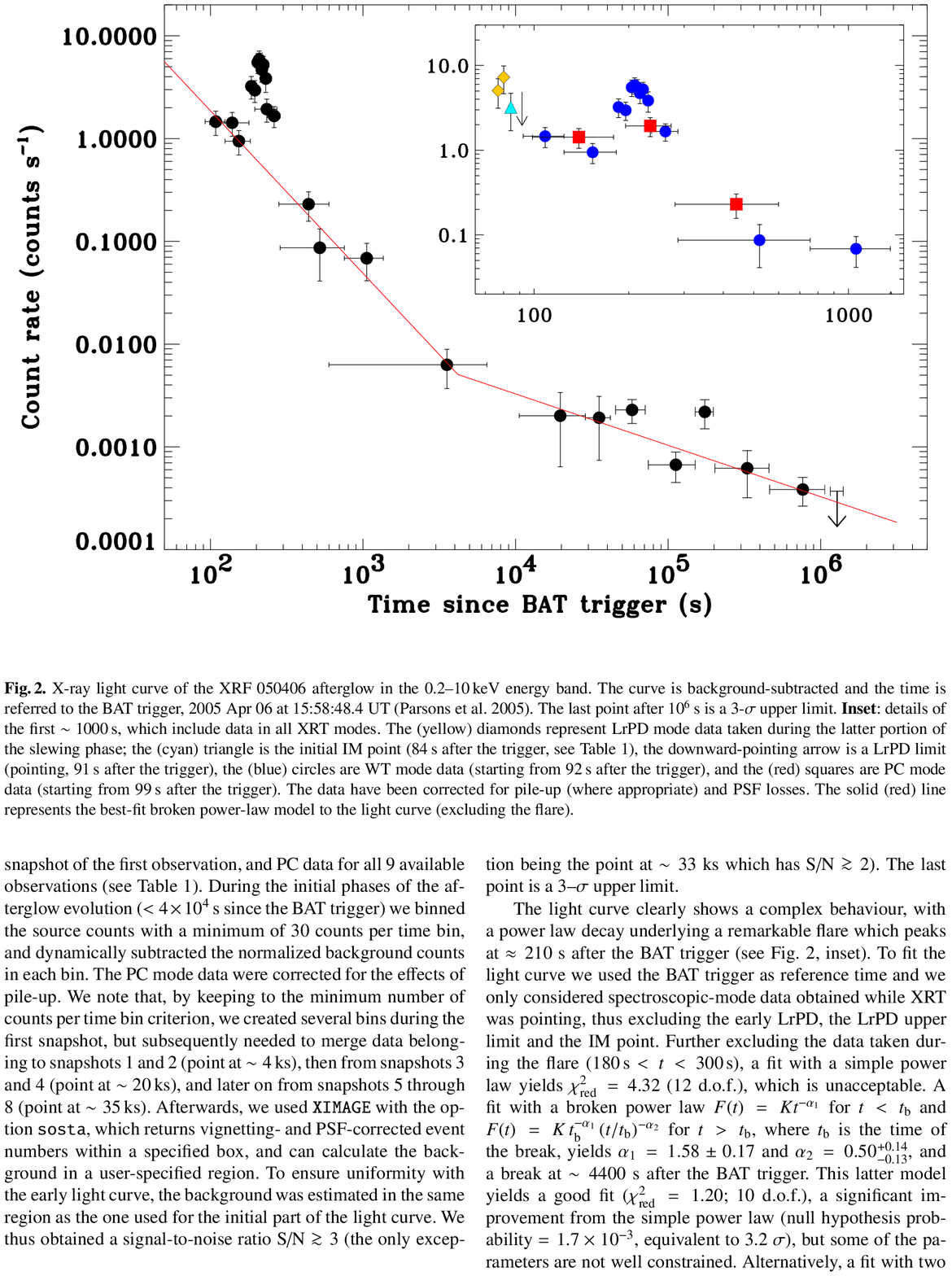}
\caption{Background-subtracted X-ray light curve of GRB~050406 (0.2-10  keV).
The red solid line shows the best-fit broken power-law model for the
points excluding the flare at about 210~s.
The inset shows details of the first 1000~s.  Yellow diamonds are data
taken in Photodiode (PD) mode, the cyan triangle is from the Image (IM) mode
frame, the blue circles are from Windowed Timing (WT) mode data, and the
red squares are Photon-Counting (PC) mode \citep[see][]{Hill2004}.
For details of the data processing and analysis, see \citet{Romano2005}.
\label{fig:050406_lc}}
\end{figure*}

Although in retrospect, the first flare observed by the XRT was
probably in GRB~050219A \citep{Tagliaferri2005, Goad2005}, the first
clear-cut example was GRB~050406 \citep{Romano2005, Burrows2005b}.
The X-ray light curve of GRB~050406 is shown in
Figure~\ref{fig:050406_lc}.
There is a strong flare peaking at about 210~s after the BAT trigger,
which rises above the underlying power-law decay by a factor of 6.
The rapid power-law decay in the first 1000~s has a decay index of
$1.58 \pm 0.17$.  At about 4400~s the light curve breaks to a flatter
decay index of $0.50^{+0.14}_{-0.13}$ \citep{Romano2005}.
When the underlying decay is subtracted, the flare itself peaks at
213~s and has rise/fall rates (expressed as power-law indices) of $\pm
6.8$.  The flare can also be fitted as a Gaussian, in which case the
width is $17.9^{+12.3}_{-4.6}$~s, and ${\delta t}/t_{peak} \sim 0.2
\ll 1$, where we take $\delta t$ = FWHM of the Gaussian =
$42.2^{+29.0}_{-10.8}$~s.

\begin{figure}
\centering
\includegraphics[width=0.9\linewidth,clip,bb=312 565 518 772]{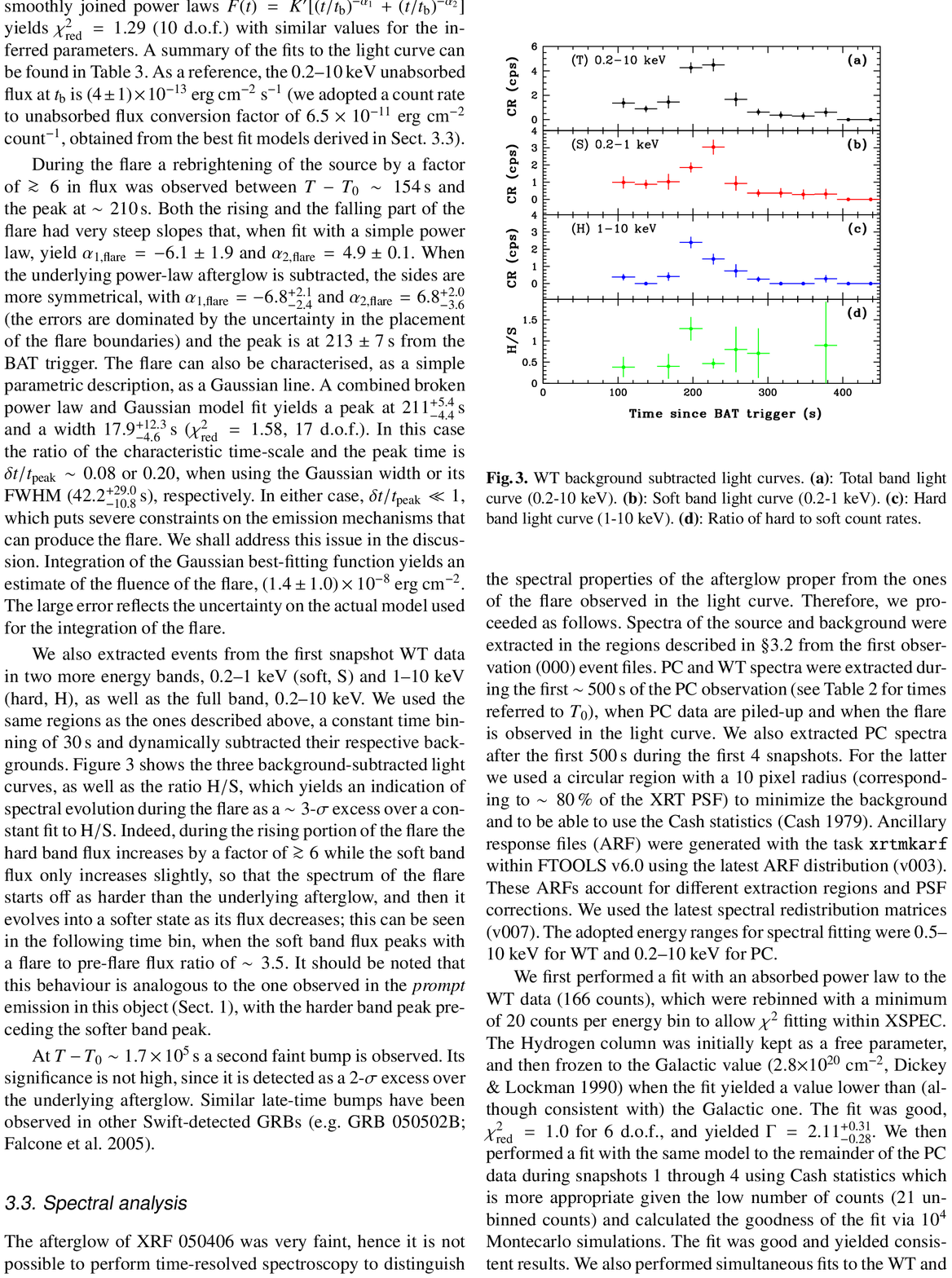}
\caption{Background-subtracted X-ray light curves of the GRB~050406 flare.
(a)~Total intensity (0.2-10 keV).  (b)~Soft band (0.2-1.0 keV).  
(c)~Hard band (1.0-10 keV).  (d)~Band ratio (H/S).
Note that the hard band peaks before the soft band; this is also
reflected in the band ratio, which is quite hard at the onset of the flare.
For details of the data processing and analysis, see \citet{Romano2005}.
\label{fig:050406_bands}}
\end{figure}

The light curve of the flare can be obtained in two energy bands to allow a search
for spectral variations.  Figure~\ref{fig:050406_bands}
shows the light curve in soft and hard bands, as well as the ratio of
the hard to soft bands.  The flare begins in the hard band, softening
significantly as it decays.  This is similar to the behavior typically
seen in the prompt gamma-ray emission from GRBs, and in particular,
seen in the prompt emission from this burst.

To reiterate, the key features seen in this flare are:
\begin{itemize}
\item Underlying afterglow consistent with a single slope before and
      after the flare.
\item Flare increases by factor of 6.
\item $\delta t/t \ll 1$ for both the rising and falling sides of the
      flare.
\item Flare softens as it progresses.
\end{itemize}

\section{GRB~050421}

\begin{figure}
\centering
\includegraphics[width=0.9\linewidth,clip,bb=132 505 483 797]{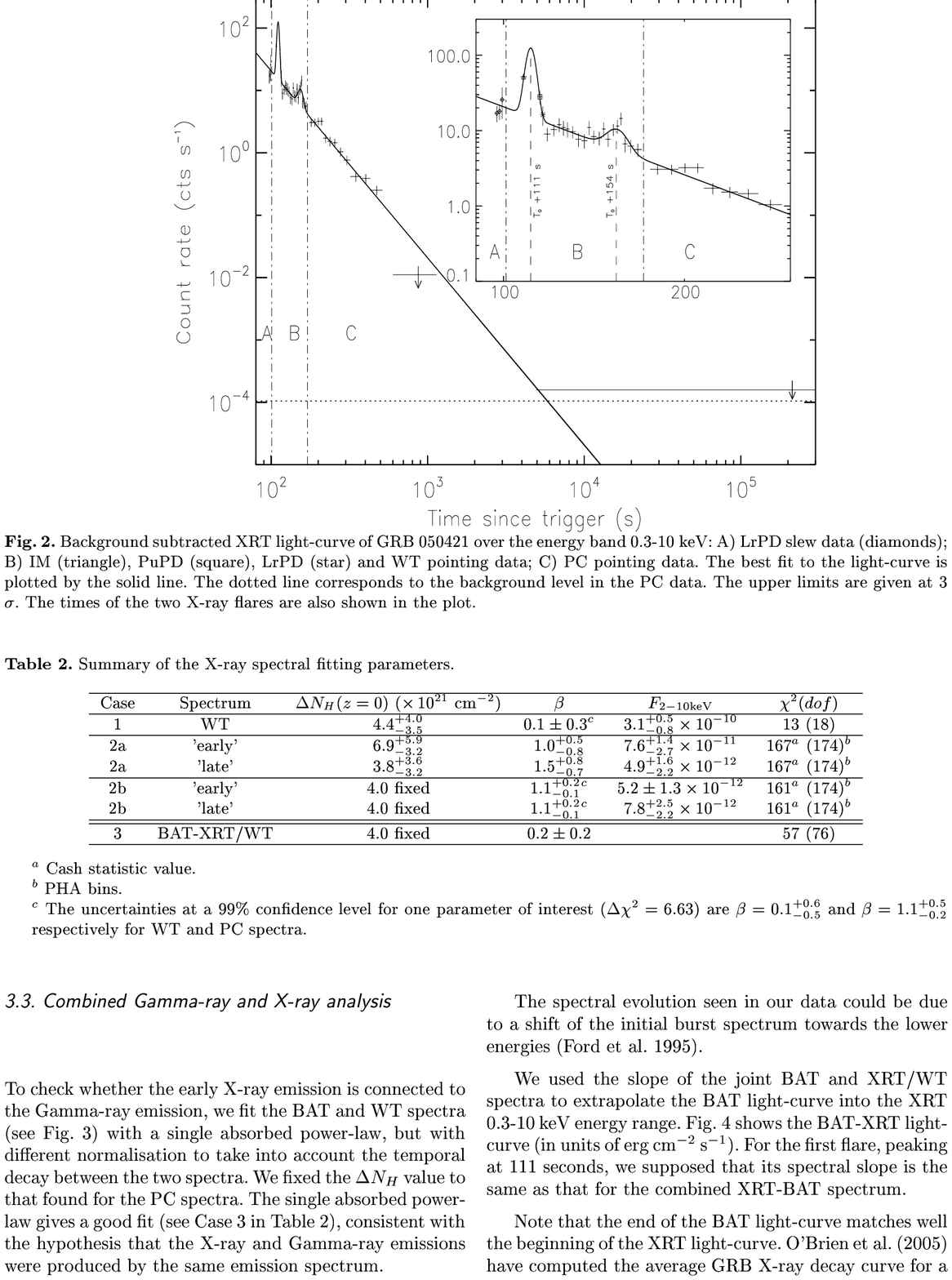}
\caption{Background-subtracted X-ray light curve of GRB~050421 (0.3-10 keV).
The solid line shows the best fit model, consisting of a power-law
plus two Gaussian flares centered at 110 and 154 seconds post-burst.
Data during time period A are in Photodiode (PD) mode (diamonds).  Time
period B has data in PD mode (square and star), and WT mode.  Data in
interval C are in PC mode.
For details of the data processing and analysis, see \citet{Godet2005}.
\label{fig:050421_lc}}
\end{figure}

The XRT observations of GRB~050421 show a large flare and a small
flare superposed on a single power-law decay with a decay index of $3.1 \pm
0.1$ (Figure~\ref{fig:050421_lc}).
Although not as well sampled as GRB 050406, primarily due to the
extremely rapid rise and fall of this flare, the X-ray light curve can
be well-modelled as a single power-law decay with two Gaussian flares superposed.
The stronger flare peaks at $111 \pm 2 s$ and has an extremely steep
rise and fall, with $\delta t/t \sim 0.07$ for both the rise and fall
times, and increases by a factor of 4 above the underlying power-law
decay \citep{Godet2005}.

Salient features of GRB~050421 include:
\begin{itemize}
\item Underlying afterglow consistent with a single slope before and
      after the flare.
\item Flare increases by factor of $\sim 4$.
\item $\delta t/t \ll 1$ for both the rising and falling sides of the
      flare.
\end{itemize}

\section{GRB~050502B}

\begin{figure}
\centering
\includegraphics[width=0.9\linewidth,clip,bb=310 536 526 710]{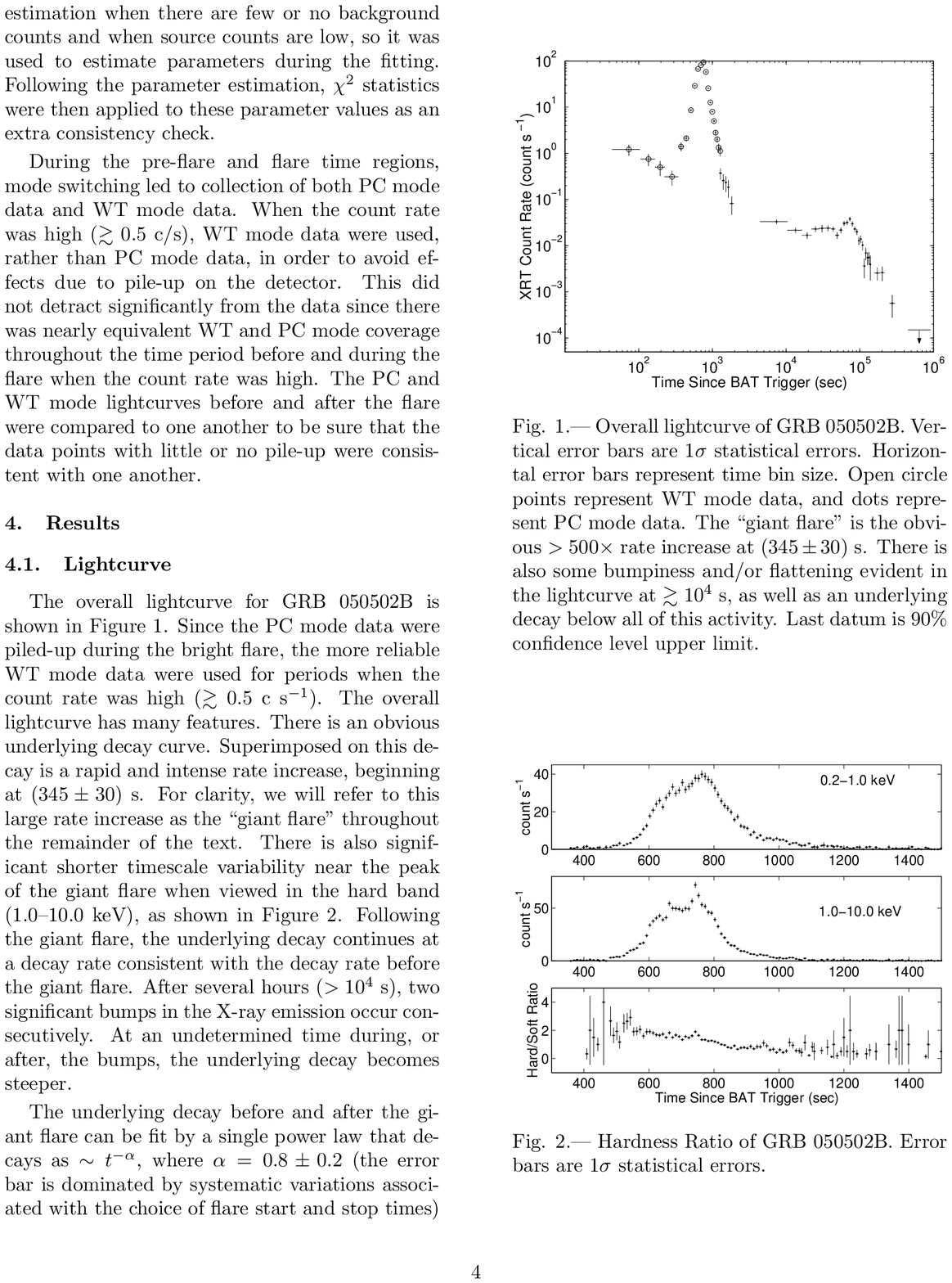}
\caption{Background-subtracted X-ray light curve of GRB~050502B (0.2-10 keV).
Open circles represent WT mode data and dots represent PC mode data.  
The latter were corrected for pile-up where necessary.
For details of the data processing and analysis, see \citet{Falcone2005}.
\label{fig:050502B_lc}}
\end{figure}

\begin{figure}
\centering
\includegraphics[width=0.9\linewidth,clip,bb=56 403 273 575]{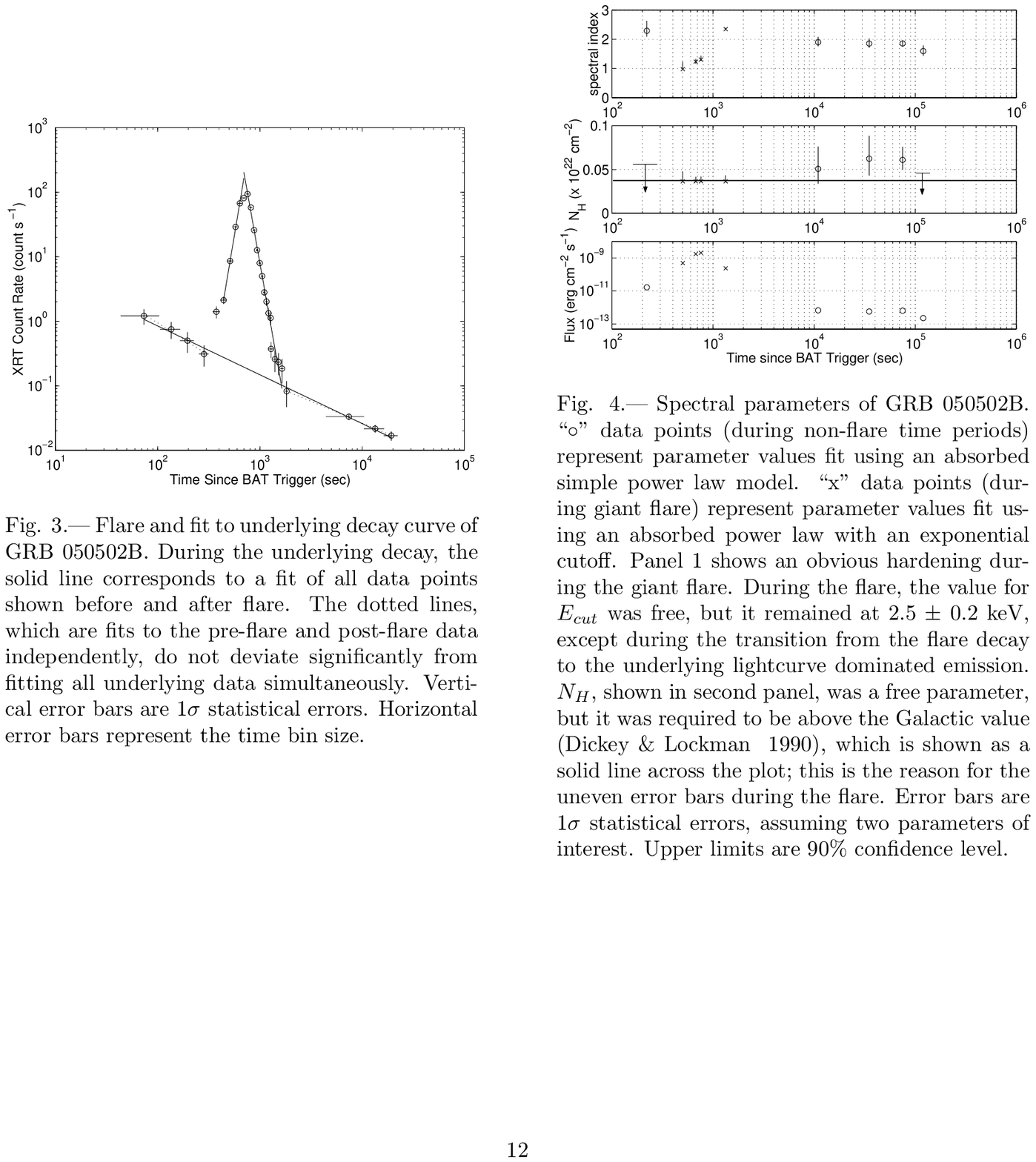}
\caption{Background-subtracted X-ray light curve of the GRB~050502B
  flare (0.2-10 keV).
The solid lines indicate the underlying afterglow (decay index of 0.8)
and fits to the flare rise and decay.
For details of the data processing and analysis, see \citet{Falcone2005}.
\label{fig:050502B_flare}}
\end{figure}

The largest flare seen to date occurred in the light curve of
GRB~050502B \citep{Falcone2005, Burrows2005b}.  The X-ray light curve
of this burst is shown in Figure~\ref{fig:050502B_lc}.
The light curve has an enormous flare peaking at about 650~s
post-burst, with late-time bumps at about 30,000~s and 700,000~s.
The main flare is shown in more detail in
Figure~\ref{fig:050502B_flare},
where the underlying afterglow is indicated by a solid line, as are
power-law fits to the rising and falling parts of the flare, which
are extremely steep, with power-law indices of about 9.5.
The fluence of the giant X-ray flare, $\sim 9 \times 10^{-7} \rm{ergs}~
\rm{cm}^{-2}$, actually exceeds the fluence ($\sim 8 \times 10^{-7} \rm{ergs}~
\rm{cm}^{-2}$)
of the prompt gamma-ray
burst detected by the Swift Burst Alert Telescope \citep{Barthelmy2005}.

\begin{figure}
\centering
\includegraphics[width=0.9\linewidth,clip,bb=310 182 524 356]{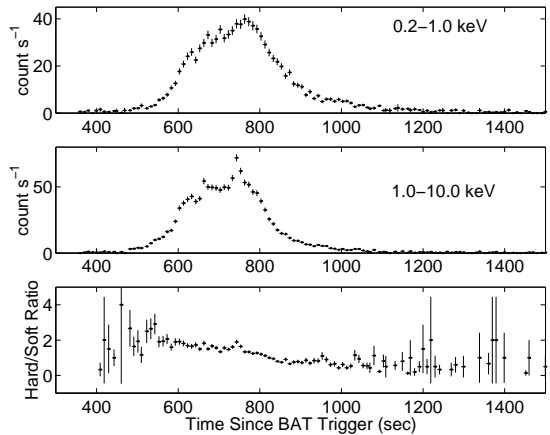}
\caption{Background-subtracted X-ray light curves of the GRB~050502B
  flare.
Top: Soft band (0.2-1.0 keV).  Middle: Hard band (1.0 - 10 keV).
Bottom: H/S band ratio.  Like GRB~050406, the beginning of the flare
is significantly harder than the preceeding afterflow, and the flare
softens as it evolves, eventually ending up with a band ratio similar
to the pre- and post-flare afterglow.
For details of the data processing and analysis, see \citet{Falcone2005}.
\label{fig:050502B_bands}}
\end{figure}

As in the case of GRB~050406, we have generated light curves in two
energy bands for this burst, shown in Figure~\ref{fig:050502B_bands}.
The flare is significantly harder than the pre- or post-flare
afterglow, and softens gradually as it evolves, with the hard band
decaying much faster than the soft band.  We note that there is
rapid variability in the hard band at about 750~s.

The key features of this flare are:
\begin{itemize}
\item Underlying afterglow consistent with a single slope before and
      after the flare.
\item Flare increases by factor of 500.
\item $\delta t/t < 1$ for both the rising and falling sides of the
      flare.
\item $\delta t/t \ll 1$ for the spike at the peak of the hard band.
\item Flare softens as it progresses.
\end{itemize}

\section{GRB~050607}
\begin{figure}
\centering
\includegraphics[width=0.9\linewidth,clip,bb=59 584 269 729]{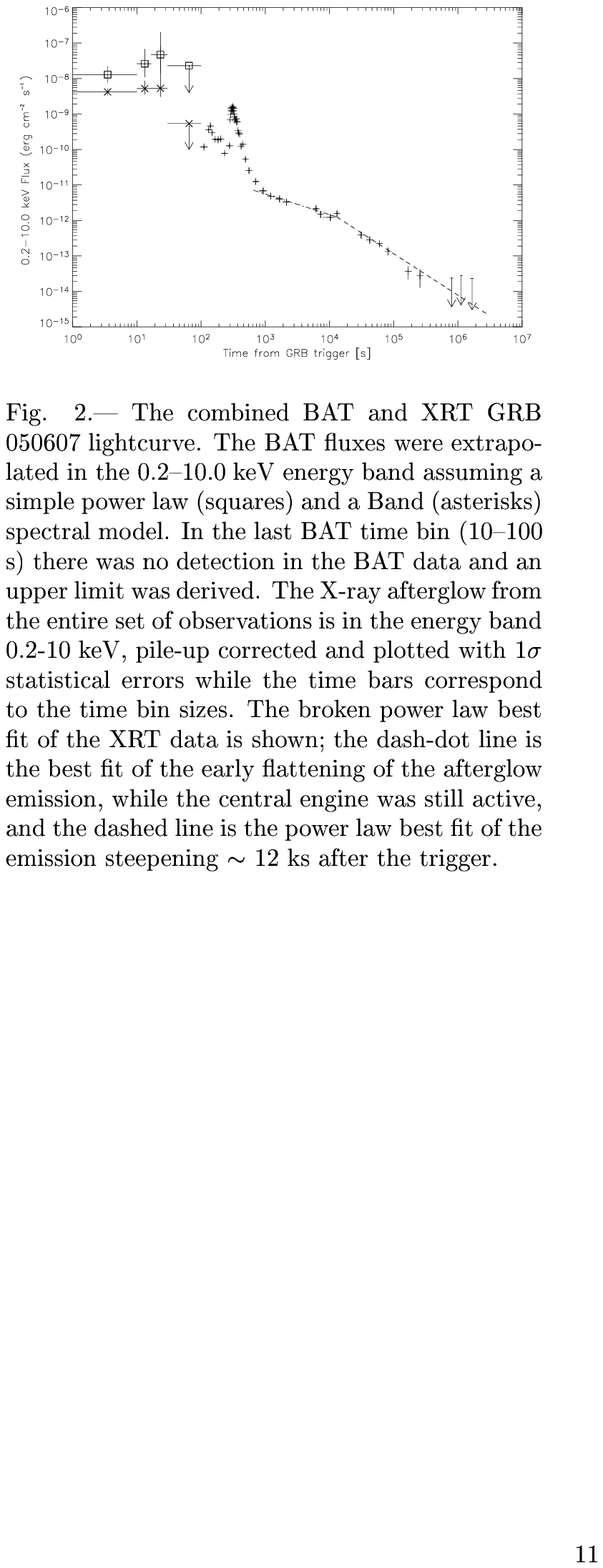}
\caption{Background-subtracted X-ray light curve of GRB~050607 (0.2-10
  keV).
This light curve includes two different extrapolations of BAT data points into the XRT
energy band (points before 100~s).  The points after 100~s are XRT
data, corrected for pile-up where necessary.
The best-fit broken power-law is shown by the dashed and dash-dot
lines.
This light curve has two early flares.
For details of the data processing and analysis, see \citet{Pagani2005}.
\label{fig:050607_lc}}
\end{figure}

The X-ray light curve of GRB~050607 is shown in
Figure~\ref{fig:050607_lc}.
Two flares are superposed on an underlying power-law decay (index
$0.58 \pm 0.07$ until about 12~ks post-burst) in the first 500~s post-burst.  In this case, we do
not measure the afterglow intensity before the first flare, which
may already be in progress when the XRT begins collecting data.  The BAT
data points have been extrapolated into the XRT band using two
different spectral models, showing that they are at least roughly
consistent with the XRT flux at about 100~s.

The main flare, which peaks at about 310~s, has a total fluence about
16\% of the BAT prompt fluence.  The rising portion of this flare is
extremely steep, with a power-law slope of about 16 when referred to
the BAT trigger, or about 2.5 when referred to the beginning of the
flare itself, and $\delta t/t \sim 0.2$.  The decay following the
flare is less steep; this flare is less symmetrical than the other
examples presented here.

\begin{figure}
\centering
\includegraphics[width=0.9\linewidth,clip,bb=59 572 272 736]{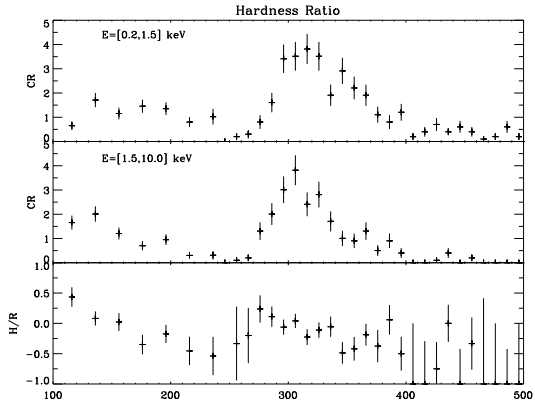}
\caption{Background-subtracted X-ray light curves of GRB~050607.
The top panel shows the soft (S) band light curve (0.2-1.5 keV in this case), the
middle panel shows the hard (H) band (1.5 - 10 keV), and the bottom panel
shows the hardness ratio, defined as (H-S)/(H+S).
For details of the data processing and analysis, see \citet{Pagani2005}.
\label{fig:050607_bands}}
\end{figure}

We have produced light curves in two energy bands to examine spectral
variations during the flare (Figure~\ref{fig:050607_bands}).
As in previous cases, the hard band rises faster than the soft band
and also decays faster, resulting in an increase in hardness ratio at
the beginning of the flare and a gradual decrease as the flare evolves
and decays.
Note that these statements are true for {\it both} flares.

Salient features of these flares include:
\begin{itemize}
\item Two flares in first 500~s.
\item Main flare increases by factor of $\sim 20$.
\item $\delta t/t \ll 1$ for the rising side of the
      flares.
\item $\delta t/t < 1$ for the falling side of the flares.
\item Both flares soften as they progress.
\end{itemize}

\section{GRB~050730}
\begin{figure}
\centering
\includegraphics[width=0.9\linewidth,clip,bb=55 260 564 672]{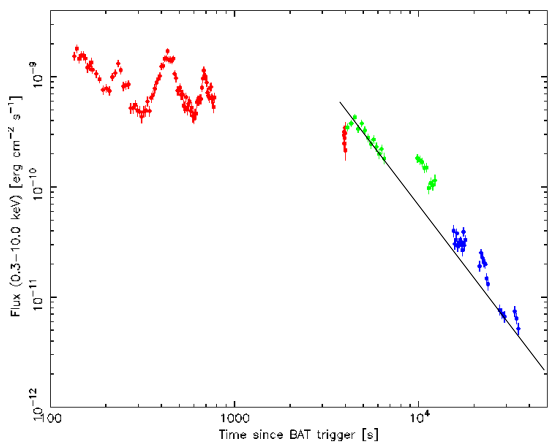}
\caption{Background-subtracted X-ray light curve of GRB~050730 (0.3-10
  keV).
A possible underlying power-law decay after the first hour is
indicated by the solid line.  However, the variability throughout this
light curve is so large that it is difficult (or impossible) to
establish the level or slope of the afterglow contribution with certainty.
\label{fig:050730_lc}}
\end{figure}

This remarkable light curve (Figure~\ref{fig:050730_lc}) shows at
least three successive flares of comparable magnitude (factor of 3--4)
and durations ($\sim 200~s$).  
Furthermore, substantial flaring and variability continue in this
light curve out to times of at least 35 ks in the observer's frame.

\section{Discussion}
The flares discussed in the preceding sections have several features
in common, which collectively point to an emission mechanism
associated with internal shocks in the GRB jet:
\begin{itemize}
\item Rapid rise and fall times, with $\delta t/t_{peak} \ll 1$.  It
      is very difficult to obtain rapid variability in the external
      shock, since the radiation physics of the shock results in 
      decay time constants no faster than $\delta t/t_{peak} \sim 1$
      \citep{Ioka2005}.  Possible mechanisms associated with the
      internal shocks and jet for these rapid variations include
      extended central engine activity, anisotropic jets, or a jet
      comprised of many ``mini-jets'' \citep{Ioka2005}.
\item Many light curves have evidence for the same afterglow intensity
      and slope before and after the flares \citep[see][for a
      counterexample]{Osborne2005}.  This argues against energy
      injection into the external shock by the flare, as would be
      expected if the flare were associated with the external shock.
\item Multiple early flares (at least 3 for GRB~050730) argue against
      one-shot explanations like the beginning of the afterglow.
\item Large flux increases (factors of tens to hundreds) are
      incompatible with origins related to the Synchrotron
      Self-Compton mechanism in the reverse shock \citep{Kobayashi2005}.
\item Flares typically soften as they progress.  This is very
      reminiscent of the behavior of the prompt emission.
\end{itemize}
Our conclusion is that the most likely explanation, which seems to
account for all of these observed features of the x-ray flares, is
that they are caused by internal shocks similar to those that produce
the prompt emission, but with lower resultant photon energies.  The
lower energies are a natural consequence of the late times (and
corresponding large radii) at which these flares are observed.

This conclusion requires extended activity by the central engine at
times long after the cessation of the prompt gamma-ray emission.  This
points in turn to a mechanism like fall-back of material into the new
black hole whose formation caused the GRB, as discussed by
\citet{MacFadyen2001} and \citet{King2005}.

More extensive discussions can be found in \citet{Nousek2005},
\citet{Zhang2005}, and \citet{Alin2005}.

\section{GRB~050904}
\begin{figure}
\centering
\includegraphics[width=0.9\linewidth,clip,bb=95 389 457 630]{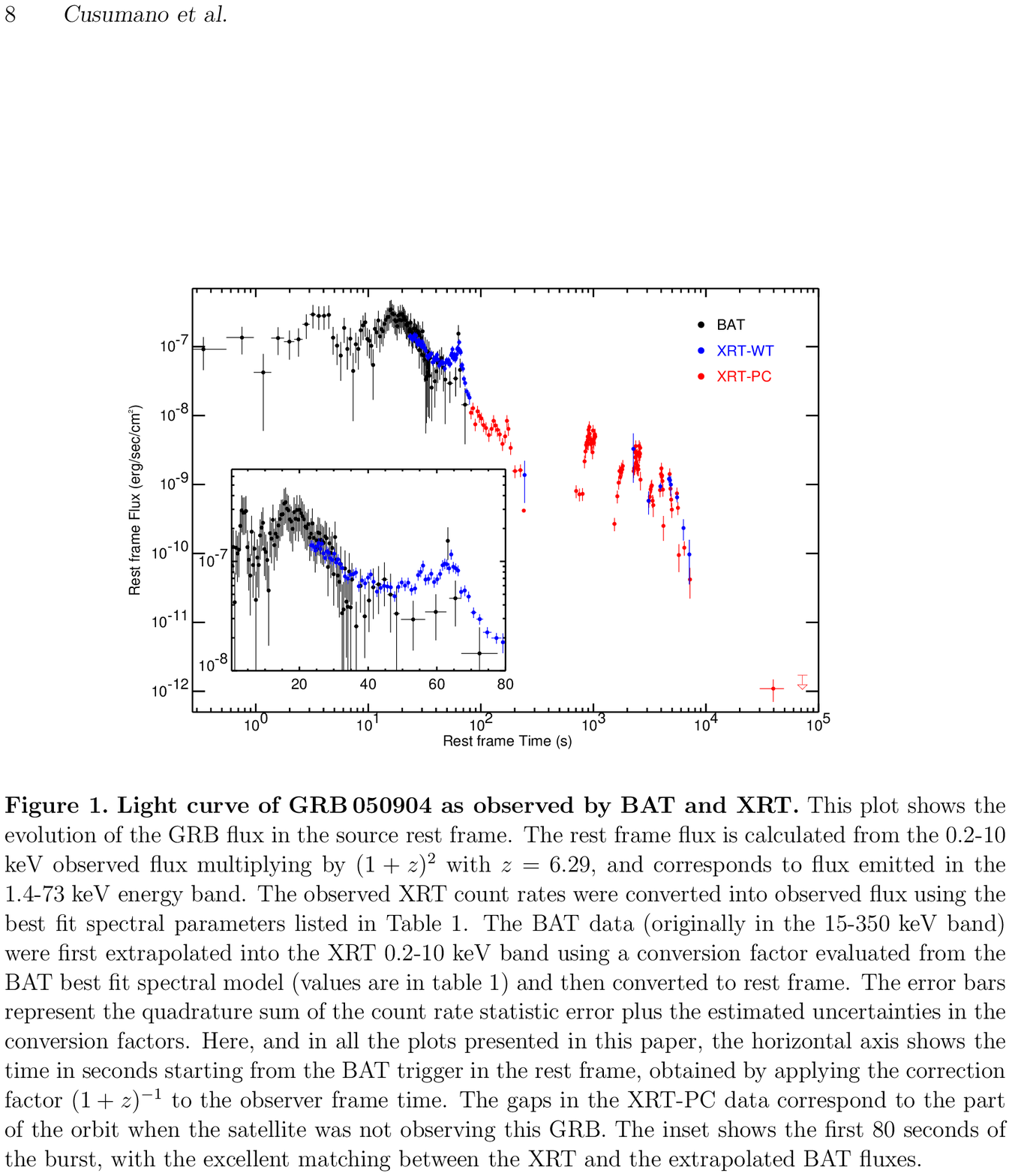}
\caption{Background-subtracted X-ray light curve of GRB~050904,
  transformed into the rest frame for z=6.29.
The plot shows the BAT light curve, extrapolated into the XRT energy
range (black points, from 0-75~s) superposed on the XRT light curve (20-10,000~s).  There is
substantial overlap in this case between the BAT and XRT data points,
which agree fairly well.
For details of the data processing and analysis, see \citet{Cusumano2005}.
\label{fig:050904_lc}}
\end{figure}

We now consider the case of GRB~050904, the highest redshift GRB
found to date.  The X-ray lightcurve of this burst is shown in
Figure~\ref{fig:050904_lc}.
The XRT light curve shows a spiky flare at about 65~s, with
substantial fluctuations in count rate extending out to nearly 10~ks
in the rest frame of the burst.

\begin{figure}
\centering
\includegraphics[width=0.9\linewidth,clip,bb=99 347 461 578]{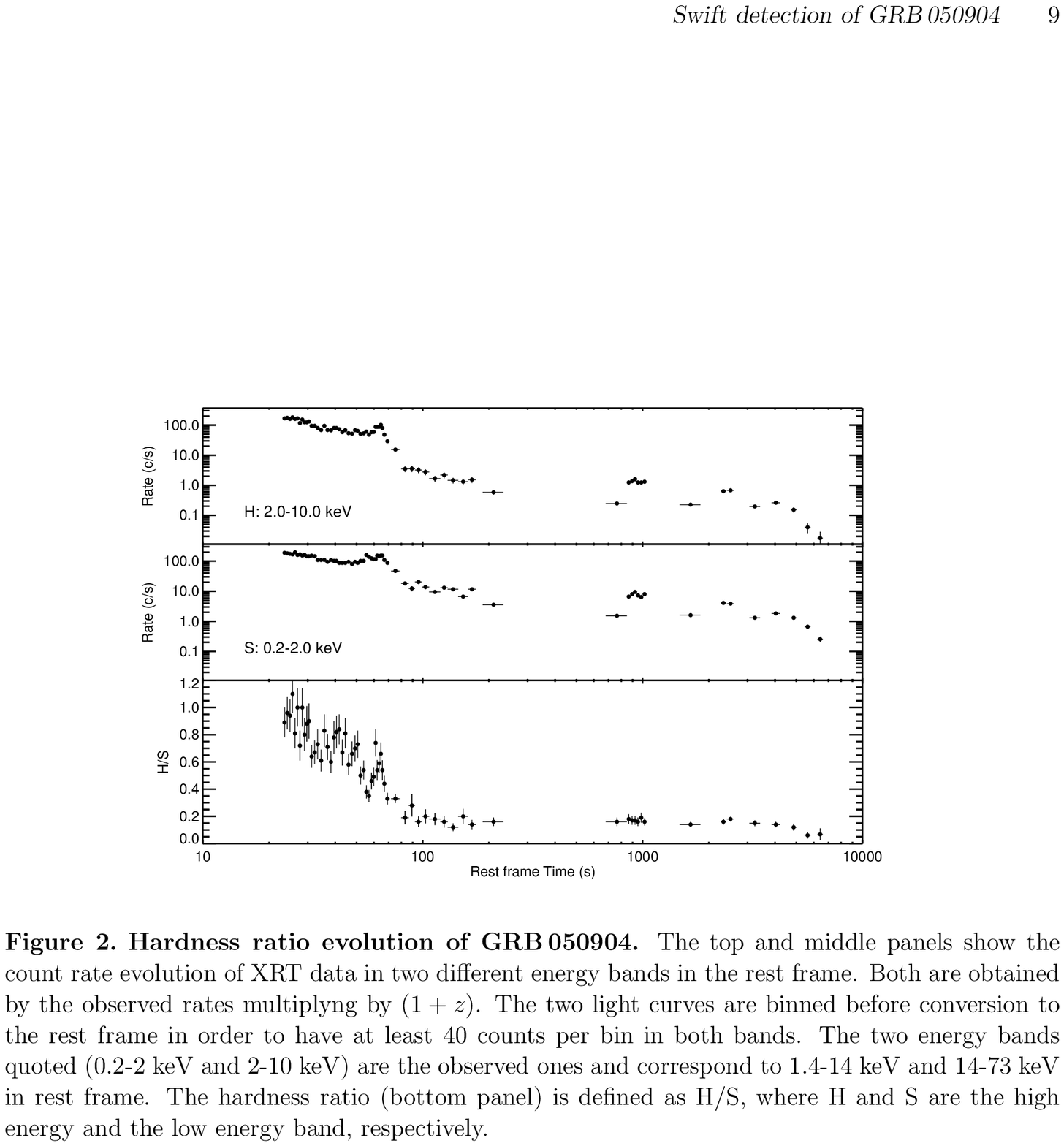}
\caption{Background-subtracted X-ray light curves of GRB~050904,
  transformed into the rest frame for z=6.29.
The upper panel shows the hard (H) band (2-10 keV in this case), the middle panel
shows the soft (S) band (0.2-2 keV), and the lower panel shows the
band ratio (H/S).
For details of the data processing and analysis, see \citet{Cusumano2005}.
\label{fig:050904_bands}}
\end{figure}

Soft and hard band light curves for GRB~050904 are shown in
Figure~\ref{fig:050904_bands}.
Unlike previous examples, the band ratio declines steadily during the
first 80~s of the XRT data, but then remains low during the late-time
variability, when the XRT count rate is varying by an order of
magnitude or more.

This light curve exhibits several differences to those discussed in
previous sections:
\begin{itemize}
\item Flaring seems to occur predominantly at much later times
      (thousands rather than hundreds of seconds).
\item The late flares exhibit no spectral variability, in stark
      contrast to the flares discussed above.
\end{itemize}
It is not clear at this time whether the differences between the
flaring activity in GRB~050904 and the bursts discussed above is
related to the high redshift of GRB~050904, although it seems quite
premature to suggest this on the basis of a single example.
Observations of additional flares, including additional high redshift
cases, will undoubtedly shed light on this over the next year.

\section{GRB~050724}

\begin{figure}
\centering
\includegraphics[width=0.9\linewidth,angle=270,scale=0.8]{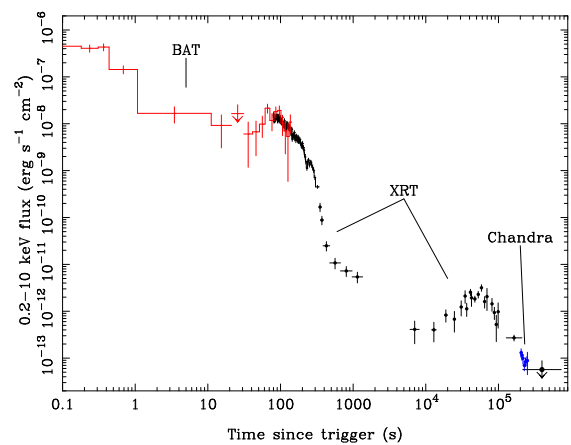}
\caption{Background-subtracted X-ray light curve of GRB~050724,
  showing BAT data extrapolated into the XRT band, XRT data, and
  Chandra data.
For details of the data processing and analysis, see \citet{Barthelmy2005b}.
\label{fig:050724_lc}}
\end{figure}

We conclude by briefly mentioning the short GRB~050724 (Figure~\ref{fig:050724_lc}).
Unlike the short GRBs 050509B \citep{Gehrels2005} and 050813
\citep{Fox2005}, which were very faint in X-rays and exhibited
power-law decays until they disappeared, the X-ray light curve of
GRB~050724 is bright, complex, and has several flare-like features.
The optical and X-ray afterglows of this burst are clearly associated
with an elliptical galaxy \citep{Barthelmy2005b}, making it very
likely that the GRB was the result of a compact object merger.  How
then, in the rarified environment of a compact binary system in the
outskirts of an elliptical galaxy, can the central engine produce
flares and bumps at the late times seen in this light curve?  One
possibility is a neutron star-black hole merger.  It is possible in
such a system, given the right initial conditions, for the black hole
to shred the neutron star, resulting in a lengthy period of in-fall
into the black hole from the shredded remnants of the neutron star.

\section{Conclusions}

Observations of GRBs by the Swift XRT have shown that flaring is very
common in X-ray light curves of GRBs and their afterglows.  It seems
likely that these flares result from extended central engine activity,
pointing toward a much longer period of activity than expected on the
basis of gamma-ray observations of prompt emission.  Further progress
in this area will undoubtedly result from statistical studies now
underway of the properties of flares as an ensemble.  Ultimately, we expect these
observational results to lead to improved theoretical models of black
hole formation and GRB central engines.

\section*{Acknowledgments}

This work is supported at Penn State by NASA contract NAS5-00136; at
the University of Leicester by the Particle Physics and Astronomy
Research Council under grant numbers PPA/G/S/00524 and PPA/Z/S/2003/00507; and at OAB by funding
from ASI under grant number I/R/039/04.

\end{document}